\documentclass[%
 reprint,
 amsmath,amssymb,
 aps,
floatfix,
]{revtex4-2}

\usepackage{graphicx}
\usepackage{braket}
\usepackage{color}
\usepackage[dvipsnames]{xcolor}
\usepackage{mathtools}
\usepackage{booktabs}
\DeclarePairedDelimiter{\ceil}{\lceil}{\rceil}

\newcommand{\figref}[1]{Fig.~\ref{fig:#1}}
\newcommand{\tblref}[1]{Table~\ref{tbl:#1}}
\newcommand{\eref}[1]{Eq.~\eqref{eq:#1}}
\newcommand{\secref}[1]{Section~\ref{sec:#1}}
\newcommand{\appref}[1]{Appendix~\ref{app:#1}}
\newcommand{\qaer}{Qiskit Aer}
\newcommand{\kyoto}{IBM Kyoto}
\newcommand{\qasm}{QasmSimulator}
\newcommand{\qr}{{Qiskit Runtime}}
\newcommand{\tn}{TensorNetwork}

\newcommand{\at}[1]{}
\newcommand{\kush}[1]{}

\begin{document}

\preprint{APS/123-QED}

\title{Qubit frugal entanglement determination with the deep multi-scale entanglement renormalization ansatz}

\author{Kushagra Garg}
\affiliation{CQST and CCNSB, International Institute of Information Technology Hyderabad, Telangana, India}
\author{Zeeshan Ahmed}
\affiliation{CQST and CCNSB, International Institute of Information Technology Hyderabad, Telangana, India}
\author{Andreas Thomasen}
\email{thomasen@qunasys.com}
\affiliation{%
QunaSys Inc. \\
Aqua Hakusan Building, 9th Floor, 1-13-7 Hakusan, Bunkyo-ku, Tokyo 113-0001, Japan
}
\date{\today}

\begin{abstract}
We study the deep multi-scale entanglement renormalization ansatz (DMERA) on quantum hardware and the causal cone of a subset of the qubits which make up the ansatz. This causal cone spans $O(M+\log{N})$ physical qubits on a quantum device, where $M$ and $N$ are the subset size and the total number qubits in the ansatz respectively. This allows for the determination of the von Neumann entanglement entropy of the $N$ qubit wave-function using $O(M+\log{N})$ qubits by diagonalization of the reduced density matrix (RDM). We show this by randomly initializing a 16-qubit DMERA and diagonalizing the resulting RDM of the $M$-qubit subsystem using density matrix simulation. As an example of practical interest, we also encode the variational ground state of the quantum critical long-range transverse field Ising model (LRTIM) on 8 spins using DMERA. We perform density matrix simulation with and without noise to obtain entanglement entropies in separate experiments using only 4 qubits. Finally we repeat the experiment on the \kyoto{} backend reproducing simulation results.
\end{abstract}

\maketitle

% Papers to cite:
% Quantum Computing Milestones: 20
% Quantum Computing Wankery industry stuff: 10
% DMERA & MERA: 10
%   MERA and CFTs: 2
%   MERA generally: 1
%   DMERA, variational algos, CFTs, noise, analytical results: 7
% NISQ: 20
%   QunaSys: 5
%   QunaSys adjacent: 2
%   Historical: 3
%   Other recent developments: 10
% Entanglement: 5
% LRTIM: 5

\section{Introduction}
Quantum systems are difficult to simulate due to their non-classical correlations arising as a consequence of quantum entanglement. However, whereas quantum entanglement poses a problem to conventional numerical methods, quantum computing promises a new paradigm where entanglement is utilized as a resource for computation \cite{ekertQuantumEntanglementQuantifiable1998, chitambarRelatingResourceTheories2016}. Quantum computers have the ability to entangle their quantum bits in a controlled fashion, thereby allowing users to encode many-body states corresponding to problems of interest. These problems range from ab-initio quantum chemistry \cite{elfvingHowWillQuantum2020, kovyrshinQuantumComputingImplementation2023, erhartConstructingLocalBases2022} and condensed matter physics \cite{cerasoliQuantumComputationSilicon2020, ohgoe2023demonstrating} to materials science \cite{delgadoSimulatingKeyProperties2022, kimFaulttolerantResourceEstimate2022} and drug discovery \cite{santagatiDrugDesignQuantum2023, bluntPerspectiveCurrentStateoftheArt2022, d.maloneSimulationLargeScale2022}. Conventionally, noisy intermediate-scale quantum (NISQ) devices have been used for problems where algorithms such as the variational quantum eigensolver \cite{VariationalQuantumEigensolver, hugginsNonorthogonalVariationalQuantum2020} (VQE) can be applied to obtain approximate solutions. Recent innovations in this area such as adapt-VQE and quantum selected configuration interaction improve the capabilities of NISQ devices, by making modifications to VQE \cite{grimsleyAdaptiveVariationalAlgorithm2019,kannoQuantumSelectedConfigurationInteraction2023a, nakagawaADAPTQSCIAdaptiveConstruction2023a}.

While entanglement can be viewed as a resource for quantum computers, by the same token the accurate evaluation of entanglement entropies is an important task. The entanglement entropy is a direct measure of how non-classical a system under study is, as well as allowing us to probe quantum phase transitions \cite{osterlohScalingEntanglementClose2002, vidalEntanglementQuantumCritical2003, chanUnitaryprojectiveEntanglementDynamics2019, leeEstimatingEntanglementEntropy2023}. A number of quantum algorithms have been developed for entanglement entropy determination \cite{wangQuantumAlgorithmsEstimating2022, laroseVariationalQuantumState2019}. Conventionally, $N$-qubit systems which display system-wide entanglement when simulated on quantum hardware using brick wall circuits require $O(N)$ depth. In the absence of error correcting codes, the total circuit depth that can be realized on a quantum computer is limited by the hardware noise experienced by the device itself. Therefore device noise becomes a limiting factor on the size of quantum systems that can be encoded on quantum hardware.

In order to circumvent this limit on circuit depth, ans\"{a}tze have been developed which are particularly suited to encode long-range entangled quantum states. One such ansatz is the deep multi-scale entanglement renormalization ansatz (DMERA) \cite{kimRobustEntanglementRenormalization2017}. DMERA has a number of attractive properties when used to encode quantum wave-functions with long-range entanglement. DMERA is the circuit implementation of the multi-scale entanglement renormalization ansatz (MERA) \cite{evenblyNonlocalScalingOperators2010} and has found similar applications. For instance, it has been used to encode renormalization group fixed points \cite{sewellPreparingRenormalizationGroup2021,aguadoEntanglementRenormalizationTopological2008,evenblyQuantumCriticalityMultiscale2013}. The depth and therefore run-time of DMERA is $\log N$, thus allowing the expression of quantum states with long-range entanglement in relatively shallow circuits. Not only does this have the potential to enable simulation of larger quantum systems, DMERA also is qubit frugal, as the causal cone of a each qubit only involves $\log N$ qubits on the quantum device. This has been shown to enable the expression of full $N$-qubit wavefunctions on devices with only $\log N$ physical qubits when mid-circuit measurement is used \cite{anandHolographicQuantumSimulation2023}. In addition, for variational quantum algorithms, it has been shown that the qMERA variant of DMERA does not have a barren plateau when used to variationally diagonalize $k\ll N$-local observables \cite{zhaoAnalyzingBarrenPlateau2021, martin2023barren}.

In this paper we study quantum states with long-range entanglement encoded using DMERA. We show that it is possible to accurately determine the entanglement entropy of the DMERA wave-function while only realizing a subset of its qubits on an actual device. To demonstrate this we simulate the variational quantum state eigensolver (VQSE) \cite{cerezoVariationalQuantumFidelity2020} both using a noisy simulator and the 127-qubit superconducting quantum processor \kyoto{}. VQSE enables the diagonalization of density matrices on quantum hardware. We here use it to obtain von Neumann entanglement entropies of 16-qubit randomly initialized DMERA. We show that this task can be accomplished while only initializing a subset of the qubits on a physical device depending on the chosen subsystem.\kush{The ref (https://arxiv.org/abs/2209.00292) using Z-X calculas proof that for a single qubit observable the qMERA doesn't have a barren plateau, and it further suggests that for k-local observable, when $k<<N$ it will still be BP free(but doesn't prove this explicitly). But they do this for a specific unitary (parameterized Rx and Rz gate on both the qubit followed by a CNOT). I was not able easily extended this for our symmetry preserving gates. They also don't consider the depth in the disentangler layer.  }\at{Thanks for clarifying, I elaborated slightly. I'd of course like to acknowledge your efforts on this more, but unfortunately I can't see a way to meaningfully add it.}

The paper is organized as follows. In \secref{methods} we review DMERA and introduce the modifications we make as compared to \cite{sewellPreparingRenormalizationGroup2021}. We here give recursive formulae showing the number of qubits that enter the causal cone of an $M$-qubit subsystem of DMERA. We then review how VQSE can be used to diagonalize the reduced density matrix of this subsystem which allows the determination of entanglement entropies. In \secref{results} we show results from applying VQSE to randomly initilialized DMERA with 16-qubits. The faithfulness of the diagonal states obtained are discussed both in the case where the subsystem is traced out from the left and from the right-hand side of the circuit, for several subsystem sizes. This is followed by a noisy simulation using device parameters obtained for \kyoto{}. In order to establish whether this methodology is applicable to cases of physical interest we then investigate the long-range interacting transverse-field Ising model (LRTIM). Using a variational ground state of LRTIM we perform noisy simulations as well as real device experiments. This is followed by \secref{conclusions} where we summarise our approach to entanglement determination in the context of NISQ and early-FTQC.

\section{Methods}
\label{sec:methods}

\subsection{DMERA}
DMERA is a quantum circuit version of the multi-scale entanglement renormalization ansatz (MERA).\kush{isn't DMERA bit different from qMERA, cause of additional parameter of depth. Should we make it more explecit.}\at{You're right, but I don't think it's necessary to make a big deal out of it in this section. In the introduction I think when talking about the ZX-calculus it makes sense to fix it though. Another difference is that qMERA structure is symmetric with respect to the qubit layout (this is not the case for DMERA, which is why the left and right-hand subsystems require different numbers of qubits to simulate)} We utilize DMERA to represent quantum states with long-range entanglement. We set the depth of each renormalization step $D=2$. For an $N$-qubit wave-function we utilize $n$ renormalization steps where
\begin{equation}
    n = \log_2{N}
\end{equation}
As with MERA, DMERA consists of isometries and disentanglers. Isometries are realized by introducing ancilla qubits at each renormalization step. An isometry is here composed of a 2-qubit gate where one input is assigned a qubit in a predetermined state. The other input is then allowed to take an arbitrary qubit as an input thus mapping it from a 2-dimensional to a 4-dimensional Hilbert space. The rest of the $D-1$ gates in this renormalization step are arranged in a brickwall pattern. In this way, the quantum circuit doubles the number of qubits at each renormalization step.
%\begin{figure}
%    \centering
%    \includegraphics[width=.2\columnwidth]{isometry.pdf}
%    \caption{A single isometry composed of a 2-qubit unitary and a single ancilla.}
%    \label{fig:isometry}
%\end{figure}
\begin{figure}
    \centering
    \includegraphics[width=.95\columnwidth]{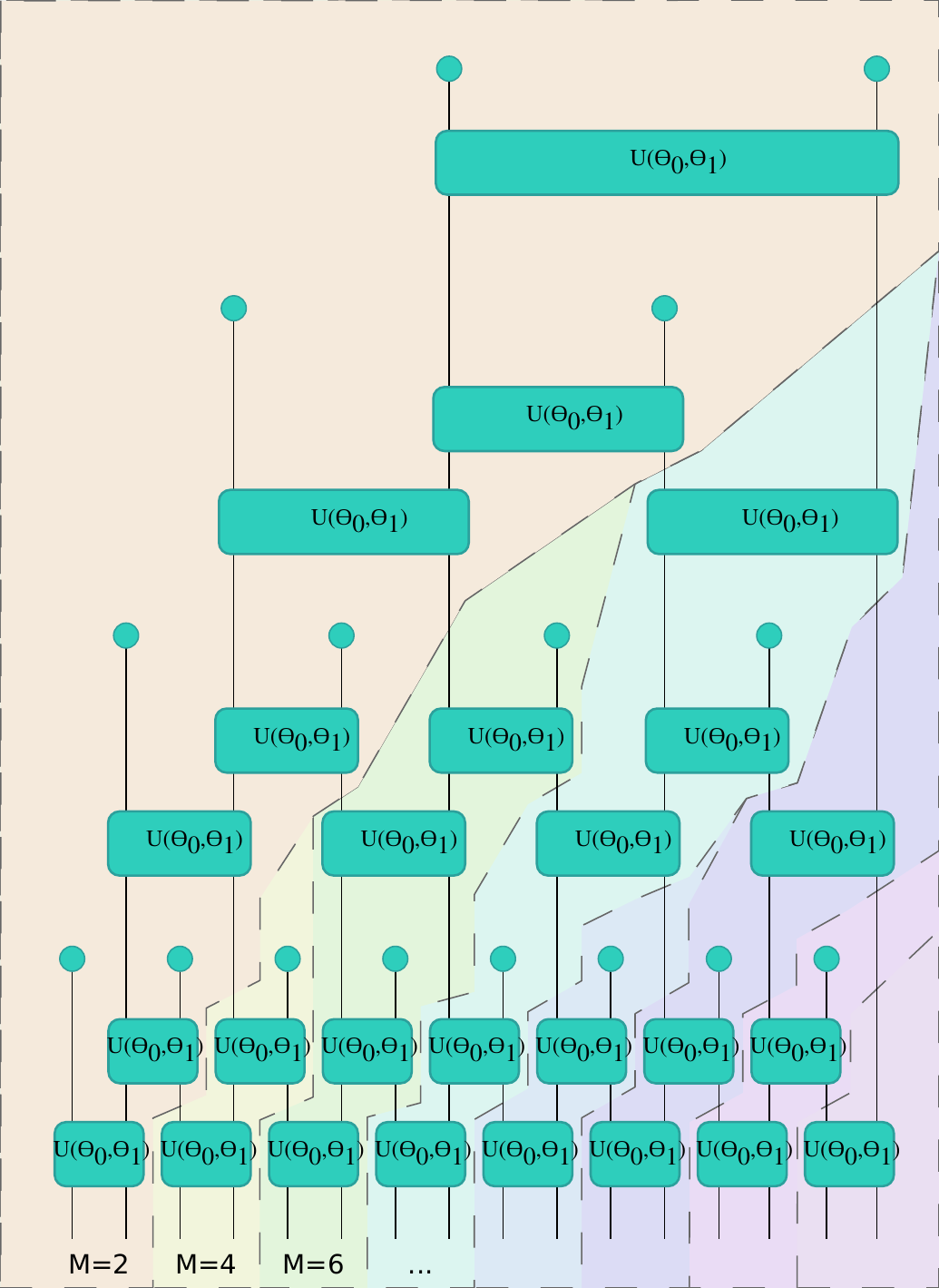}
    \caption{The structure of a 16-qubit DMERA. The shaded regions depict the causal cones of $M$-qubit subsystems spanning qubits ranging from the left-most one to the $M$'th one. Circles represent qubits.}
    \label{fig:16qbit_dmera}
\end{figure}

The structure of a 16-qubit DMERA is depicted on \figref{16qbit_dmera}. By convention we construct the isometries here by assigning the ancilla qubit to the right-hand input and routing input qubits to the left-hand input. However, the gate-fabric is chosen such that this assignment of left and right inputs is arbitrary at the level of each 2-qubit gate (see Appendix~\ref{app:circuit}).  In contrast to the original formulation of DMERA by Kim and Schwingle \cite{kimRobustEntanglementRenormalization2017} and subsequent work by Sewell and Jordan \cite{sewellPreparingRenormalizationGroup2021} the initial layer in our construction consists of only a single 2-qubit gate.

Although any subsystem chosen will have a causal cone with depth $\log{N}$, the number of physical qubits that make up the causal cone depends on the subsystem chosen. Here we will focus on two different cases, one in which the subsystem consists of the sequence of $M$ left-most qubits, and the other being the $M$ right-most ones. For the left-most case we have
\begin{equation}
    N_\textup{physical} = \log_2{N} + \sum_{i=1}^{\log_2N} \ceil{M/2^i}.
    \label{eq:n_q_l}
\end{equation}
When the subsystem is chosen from the $M$ right-most qubits we obtain
\begin{equation}
    N_\textup{physical} = \sum_{i=1}^{\log_2N} \ceil{M_i/2}
    \label{eq:n_q_r}
\end{equation}
where the $M_i$ are defined recursively as
\begin{equation}
    M_i = \ceil{M_{i-1}/2} + 1
\end{equation}
and $M_0\equiv M$. In both cases while the DMERA wave-function itself by construction posseses system-wide entanglement, only a subset of the physical qubits are needed to simulate it on quantum hardware. The number of qubits depends linearly on $M$ and has a logarithmic dependence on $N$.
\begin{table}
\caption{\label{tbl:physical}Number of physical qubits inside the causal cone of an $M$-qubit subsystem belonging to an $N$-qubit DMERA.}
\begin{tabular}{ccccccc}
\toprule
$N$ & $D_{\textup{ECR-gates}}$ & $M$ & $N_{\textup{Physical}, L}$ & $N_{\textup{Physical}, R}$  & $N_{\textup{Angles}, L}$ & $N_{\textup{Angles}, R}$ \\
\midrule
$16$ & $14$ & $2$ & $8$ & $5$ & $14$ & $14$ \\
$16$ & $14$ & $3$-$4$ & $9$ & $8$ & $18$ & $24$ \\
$16$ & $14$ & $5$-$6$ & $11$ & $9$ & $26$ & $28$ \\
$8$ & $10$ & $2$ & $6$ & $4$ & $10$ & $10$ \\
$8$ & $10$ & $3$-$4$ & $7$ & $6$ & $14$ & $18$ \\
\bottomrule
\end{tabular}
\end{table}

We choose a parity preserving gate-fabric of the form
\begin{equation}
\label{eq:gate-fabric}
    U(\theta_0,\theta_1) = \begin{bmatrix}
        \cos \theta_1 & 0 & 0 & -\sin \theta_1 \\
        0 & \cos \theta_0 & -\sin \theta_0 & 0 \\
        0 & \sin \theta_0 & \cos \theta_0 & 0 \\
        \sin \theta_1 & 0 & 0 & \cos \theta_1
    \end{bmatrix}
\end{equation}
This can be decomposed as
\begin{equation}
    U(\theta_0,\theta_1) = R_{xx}(\pi/2)(R_z(\phi_1)\otimes R_z(\phi_0))R_{xx}(-\pi/2)
\end{equation}
where $\phi_0 = -\theta_0 + \theta_1$ and $\phi_1 = \theta_0 + \theta_1$. The full wave-function resulting from a DMERA with this gate-fabric then has a parity which is determined by the initialization of the ancilla qubits. In each renormalization layer qubits were alternatingly initialized as $\ket{0}$ and $\ket{1}$, which for $N = 2^{n}$ results in even parity when $n \geq 2$.

We summarize the physical qubit counts relevant to the circuits we use in this work in \tblref{physical}. Here we also list the depth of the circuits in terms of the number of ECR-gates obtained from circuit transpilation into an OpenQASM compatible gate-set using \qr{}. Also listed are the numbers of gate-angles contained in each circuit. The general tendency is for the subsystem defined by the left-most qubits of equal $M$ to the corresponding right-most one to have more qubits, but fewer isometries and disentanglers and therefore gate-angles.

\subsection{Variational Quantum State Eigensolver}
In this section we introduce the Variational Quantum State Eigensolver (VQSE), which is the method by which we determine entanglement entropies. The mechanism by which this is achieved is to variationally diagonalize the RDM of a subsystem of interest. In the following we summarize the flow of the algorithm and explain how it is used in this paper.
\begin{figure*}[ht!]
    \centering
    \includegraphics[width=.7\textwidth]{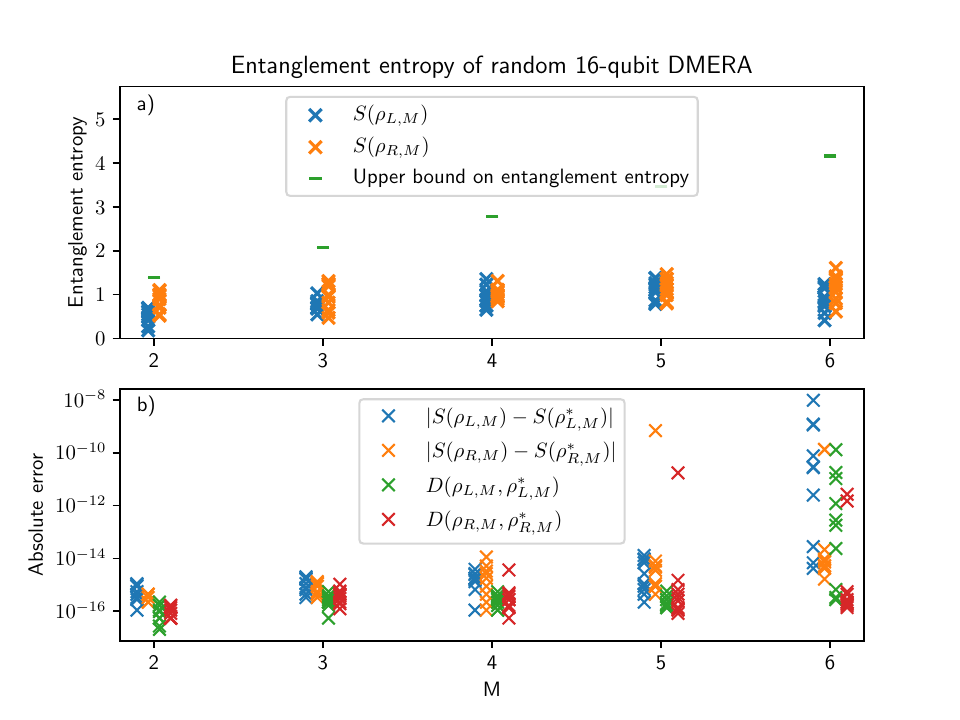}
    \caption{Density-matrix simulation results for randomly initialized DMERA diagonalized using VQSE. (a) The exact entanglement entropy obtained as a function of subsystem size for the subsystem taken from the left (blue crosses) and from the right (orange crosses). The upper bound on the entanglement entropy for each given subsystem size is indicated (green bars). (b) The absolute error in entanglement entropy found by VQSE is plotted (blue and orange crosses) along with the trace distance from the exactly diagonalized density matrix to the one estimated by VQSE. Each datapoint represents a DMERA randomized with a unique seed. For each $M$ the experiment has been run 10 times. The data has been shifted slightly away from integer values of $M$ for better visual clarity.}
    \label{fig:random_dmera}
\end{figure*}

Tracing out a subsystem of a pure quantum wave-function leaves behind the mixed state of the remaining $b$ subsystem. The RDM can be written in the form of its eigen-decomposition as
\begin{equation}
    \rho_b = \sum_i \lambda_i \ket{e_i}\bra{e_i}.
\end{equation}
Without loss of generality we order the $\lambda_i$ as
\begin{equation}
    \lambda_0 \geq \lambda_1 \geq \ldots 
\end{equation}
If this diagonalization is known, non-linear functions of the RDM can be easily calculated. As an example, we consider the von Neumann entropy which is
\begin{align}
    S &= -\textup{tr} \bigl(\rho_b \ln \rho_b \bigr) \\
      &= - \sum_i\lambda_i \ln \lambda_i.
\end{align}
The VQSE evaluates the von Neumann entanglement entropy and possible other quantities that require a diagonalization of the RDM by performing a variational optimization, which brings $\rho_b$ into a known basis. We consider the Hamiltonian
\begin{equation}
    H =  \sum_i\epsilon_i \ket{i}\bra{i}
\end{equation}
where $\ket{i}$ are computational basis states. $\epsilon_i$ are chosen such that
\begin{equation}
    \epsilon_0 < \epsilon_1 < \ldots.
\end{equation}
We then define the loss function
\begin{align}
    \mathcal{L}_\epsilon(\theta) &= \textup{tr}\bigl(H U(\theta) \rho_b U^\dagger(\theta) \bigr) \\
    &=  \sum_i\textup{tr}\bigl(\epsilon_i \ket{i}\bra{i} U(\theta) \rho_b U^\dagger(\theta) \bigr).
\end{align}
The global minimum of the loss function is achieved when
\begin{align}
    \rho_b^* &= U(\theta^*) \rho_b U^\dagger(\theta^*) \\
    &=  \sum_i\lambda_i U(\theta^*) \ket{e_i}\bra{e_i} U^\dagger(\theta^*) \\
    &=  \sum_i\lambda_i \ket{i}\bra{i}.
\end{align}
The unitary $U(\theta^*)$ diagonalizes the RDM by rotating each of its eigenstates into the computational basis, which is the eigen-basis of the Hamiltonian. This results in
\begin{equation}
    \mathcal{L}_\epsilon(\theta^*) =  \sum_i\epsilon_i \lambda_i.
\end{equation}

After the variational optimization converges, $\lambda_i$ can be obtained by sampling $\rho_b^*$ in the computational basis. The Hamiltonian chosen for VQSE does not need to be diagonal in the computational basis, however, the immediate benefit is that the loss function will be measurable without added depth from measurement circuits.

In this study we evaluate the expectation values of Kraus operators corresponding to each diagonal density matrix element. In this way we are able to use linear zero noise extrapolation of \qr{}, however the returned $\lambda_i$ are not guaranteed to be positive. For this reason we truncate their value to $0$. We refer to \appref{circuit} for the details of the circuits used for VQSE and the circuit routing.
\begin{figure*}
    \centering
    \includegraphics[width=.7\textwidth]{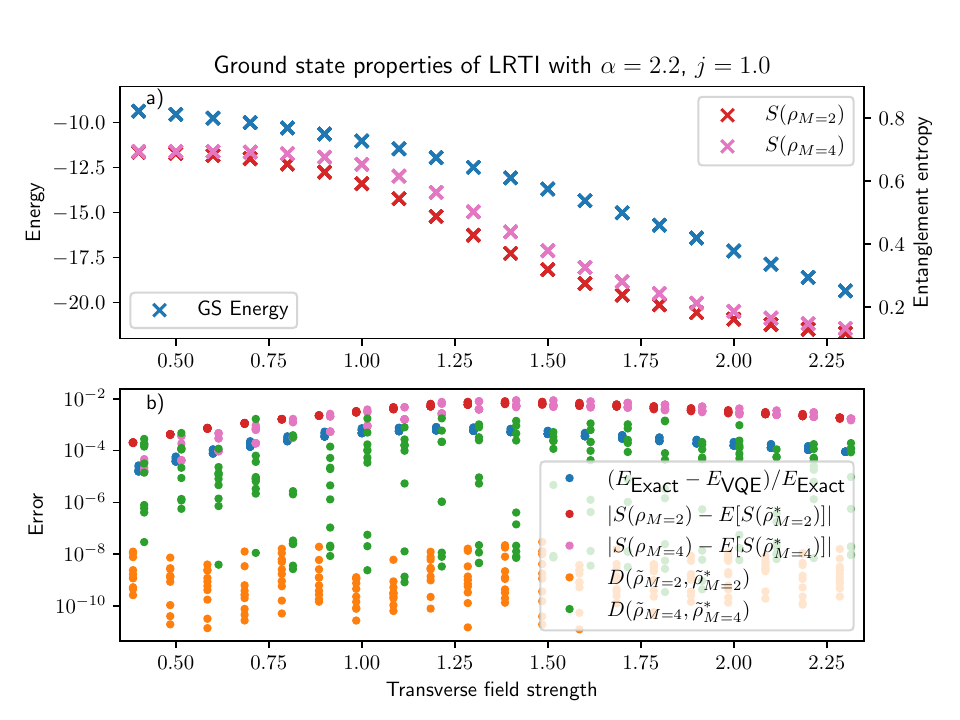}
    \caption{VQE and VQSE results for the LRTIM using state-vector and density-matrix simulation. (a) The exact ground state energy (blue crosses), entanglement entropy for $M=2$ (red crosses) and that for $M=4$ (pink crosses) are plotted as a function of the transverse field. (b) Also plotted is the error in the variational ground state energy relative to the exact ground state energy (blue dots), the absolute error of the VQSE entanglement entropy of the variational ground state both for $M=2$ (red dots) and  $M=4$ (pink dots). We also plot the trace distance between the exactly diagonalized density matrix of the $M=2$ (orange dots) and $M=4$ (green dots) subsystem of the VQE ground state and the corresponding VQSE density matrix. For visual clarity the datapoints are shifted slightly on the first axis. For each value of the transverse field strength the experiment has been run 10 times.}
    \label{fig:lrti_real}
\end{figure*}

\section{Results}
\label{sec:results}
We study the entanglement entropy obtained by variational diagonalization of an $M$-qubit subsystem of DMERA. Initially we consider whether the entanglement entropy of an arbitrary DMERA wave-function can be accurately assessed by VQSE. Then we consider the physically motivated case of the long-range interacting transverse field Ising model (LRTIM). We first verify that DMERA can accurately represent the ground state of LRTIM undergoing a quantum phase transition. We then obtain the entanglement entropy via state-vector simulation as well as noisy simulation and real device experiments.

\subsection{Random DMERA}
We randomly initialize a $N=16$ DMERA and perform VQSE. We compare the entanglement entropy found by VQSE to the exact one for a range of different subsystem sizes. We also calculate the trace distance between the VQSE output density matrix and the exact diagonal density matrix.

On \figref{random_dmera} the notation $\rho_{L/R,M}$ describes the exactly diagonalized $M$-qubit subsystem density matrix where the qubits are taken to be the leftmost/rightmost qubits. $\rho_{L/R,M}^*$ is the corresponding variationally diagonalized density matrix, $S(\cdot)$ is the von Neumann entropy and $D(\cdot)$ is the trace distance. The entanglement entropy is exact to within about $10^{-8}$ in the worst case with $M=6$. The validity of the solution found is supported by the trace distance, which shows similar magnitude.

We thereby prove the faithfulness of the density matrix obtained by VQSE for DMERA. We remark here that the quantum circuits simulated when obtaining the $M$-qubit subsystem contained only the number of qubits described by \eref{n_q_l} and in \eref{n_q_r}. For instance the entanglement entropy of the $M=6$ subsystem was obtained using only $11$ physical qubits. As seen on the figure, there is no appreciable difference between the results obtained when the subsystem is taken from the left as compared to the right. This is interesting as there is a significant difference in the number of physical qubits that enter each circuit simulating the wave-function, but the entanglement entropy is not significantly different.

\subsection{LRTIM}
We study the LRTIM with ferromagnetic interaction, whose Hamiltonian is
\begin{equation}
    H = h\sum_{i} S_i^z - J\sum_{i<j} \frac{1}{|j - i|^{-\alpha}} S_i^xS_j^x.
\end{equation}
Setting $\alpha = 2.2$ we obtain the ground state entanglement entropy using VQSE in the range $0.4J < h < 2.3J$ for $N = 8$. We perform the density matrix diagonalization with $M = 2$ and $M = 4$. For these experiments we first performed VQE using state-vector simulation to obtain the DMERA which most faithfully reproduces the ground state energy in the given parameter range. We then traced out the $M=2$ and $M=4$ subssystems and performed the state diagonalization to obtain entanglement entropies. This is done using first using density-matrix simulation without noise to asses the accuracy of the diagonal RDM obtained by VQSE. In the section thereafter we redo our numerical experiments using noise.

\begin{figure*}[t!]
    \centering
    \includegraphics[width=0.49\textwidth, trim={20px 0 30px 0}]{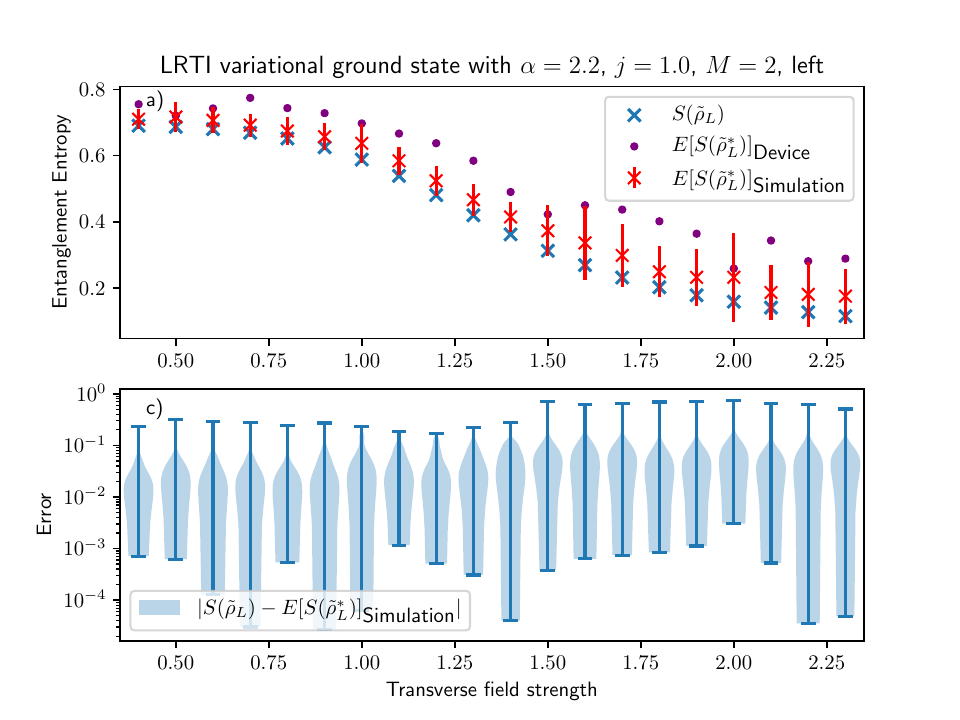}
    \includegraphics[width=0.49\textwidth, trim={20px 0 30px 0}]{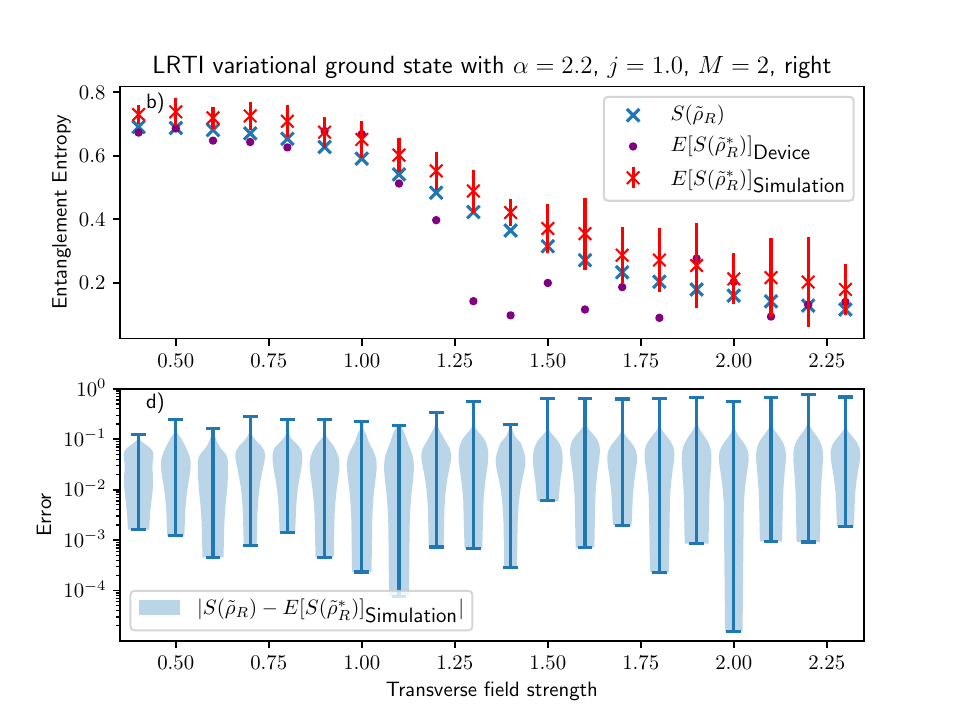}
    \caption{Results from VQSE using the noisy simulator and real device for $M=2$ both for the left-hand subsystem (left figure) and the right-hand one (right figure). (a-b) The mean entanglement entropy obtained from the noise simulator (red crosses) and real device (purple dot) are depicted along the exact entanglement entropy (blue crosses). The standard deviation of the noise simulator result, indicated using error bars, is a sample estimate over the 100 repetitions of the simulation. (c-d) The absolute error of the entanglement entropy obtained as a sample estimate with the noise simulator. The violin plot shows the best and worst outcomes and interpolates a probability density based on 100 repeated experiments.}
    \label{fig:qiskit_noisy}
\end{figure*}

\subsubsection{Density-matrix Simulation}
On \figref{lrti_real} we examine how accurately the RDM was diagonalized by VQSE. The RDM of the exact ground state is denoted $\rho_M$ while the RDM of the VQE ground state is $\tilde{\rho}_M$ and the diagonal RDM found by VQSE is $\tilde{\rho}_M^{*}$. We denote the estimate of the von Neumann entropy obtained by evaluating the diagonal entries of the RDM as $E[S(\cdot)]$. The variational ground state energies are accurate to an error of about $10^{-4}$ to $10^{-3}$, whereas the error in the entanglement entropy is $10^{-3}$ to $10^{-2}$. We also calculated the trace distance between the VQSE diagonal state and the exactly diagonalized RDM of the variational ground state. The trace distance tends to be $10^{-11}$ to $10^{-8}$ for $M=2$, whereas it lies in the range $10^{-8}$ to $10^{-3}$ for $M=4$. For $M=2$ this result suggests that the error in the entanglement entropy mainly arises from the variational ground state found by VQE and not from VQSE as VQSE can reproduce the exact diagonal with high accuracy. This also explains how for the $M=4$ case the entanglement entropy found is closer to the exact value, even though VQSE results in a worse diagonal RDM.

DMRG studies of LRTIM \cite{zhuFidelityCriticalityQuantum2018} show that in the thermodynamic limit the critical field is given by $h_c\approx 2.1 J$. With a system size of just 8 qubits we cannot extrapolate towards the thermodynamic limit, but our result is consistent with the expected volume law entanglement entropy. It appears that DMERA can reproduce the ground state energy of LRTIM to a relative accuracy of about $10^{-4}$ to $10^{-3}$ and VQSE can then be utilized to diagonalize the $M=2$ and $M=4$ subsystems. However, the variational DMERA ground state does not appear to reproduce the entanglement entropy as accurately.

\subsubsection{Noise simulator and real device results}
In this section we show results from simulation with a noise model and real device experimentation. The noise simulation was carried out using \qaer{} using calibration data from \kyoto{}. Although we were not able to perform the density matrix diagonalization on \kyoto{} itself, using \qaer{} we scheduled 100 simulations for each value of $h$. Each batch of simulations are based on a single variational ground state of the LRTIM.

\figref{qiskit_noisy} shows the results both from the noise simulator and the real device. The exact ground state entanglement entropy is shown along with results from 100 randomly seeded runs of VQSE using \qaer{}. All results were obtained using output measurement error mitigation as well as linear zero noise extrapolation. The real device results were obtained by first performing the density matrix diagonalization using the \qasm{}, then after finishing the diagonalization, the variationally optimized gate-angles were used to obtain the entanglement entropy on \kyoto{}. One additional seed distinct from those used with the noise simulator was used in the final simulation using the \qasm{}.

Contrasting the results from the noisy experiments to the density matrix simulation, we see that in addition to the statistical uncertainty caused by noise, there is also a systematic error, which causes the estimated entanglement entropy to be higher in general than it otherwise would be. This is likely due to the tendency of noise to introduce randomness to the state, which gives it inherently higher entropy. In addition to this, measurement output noise directly impacts the estimation in such a way that the state will seem to have a higher entropy.

The results also show large deviations appearing on \figref{qiskit_noisy}(b). We speculate that this may be caused by drift of device parameters, which may not be reflected in the calibration data used by the noise model. This has the additional effect that the effectiveness of error mitigation may change in real time - an effect which is not taken into account by the noise modeling.

The mean absolute error in entanglement entropy as a result of noise ranges from $0.02 - 0.08$. To our knowledge this is the first time entanglement entropies of an at least $8$-qubit wave-function have been obtained with VQSE in the presence of noise. Our ability to carry out this calculation relies on the ability of DMERA to express entanglement of $N$ qubits with $\log(N)$ depth circuits using $\log(N) + M$ physical qubits. It is remarkable that the full VQSE diagonalization can be carried out in the presence of noise. Due to the practical limitations of executing quantum circuits on real devices, we have not had the time to do the full diagonalization on one. Our results suggest that it could be done given enough resources.

\section{Conclusions}
\label{sec:conclusions}
Quantum circuits can be used to express long-range entanglement between qubits in order to simulate physical systems of interest. Quantifying the entanglement by interrogating a partition of the entire system requires all quantum bits and instructions in the causal cone of that partition to be included in a circuit that simulates the full wave-function.

We have used DMERA as an ansatz, which enabled us to construct a causal cone with $\mathcal{O}(M + \log N)$ physical qubits. Doing so for randomly initialized DMERA with 16 qubits we were able to show that the reduced density matrix can be diagonalized with the right choice of ansatz.

We then showed using density-matrix simulation that for a system of physical interest, the LRTIM, the entanglement entropies can accurately be reproduced at parameter values that are consistent with a quantum phase transition. This was repeated using a noise simulator, where even in the presence of noise the results from VQSE on average approach the exact results.

Finally we also evaluated the entanglement entropy of the VQSE RDM by using measurement data from \kyoto{}. On the real device, we see some deviation from the results obtained from our simulation, which hints that noise characteristics of the device in real time, may deviate from reported calibration values due to device drift, however the result otherwise agrees well with simulations.

It is remarkable that it is possible to determine the entanglement entropy of an 8-qubit system using VQSE. This is enabled by the shallow circuits realized by DMERA and by involving as few qubits as possible in the simulation of the full wave-function. It seems feasible to extend this scheme to demonstrate entanglement entropies of much larger circuits as the resources required for the entanglement determination scales as $\log{N}$. We did not experiment with $M>2$ due to the additional layers needed in the VQSE circuit. The diagonalization circuit used in VQSE was chosen as a brickwall circuit, however, it may also be possible to use DMERA or similar entanglement renormalization circuits to reduce the depth of the diagonalization circuit. As we have seen, the main error observed in the entanglement entropy arose from the variational ground state chosen for the LRTIM, not from VQSE itself. In conclusion there appears to be room for experimentation with shallower VQSE circuits with potentially $\log{M}$ depth scaling.

\begin{acknowledgements}
We thank Y. O. Nakagawa for discussions during the early part of this work. We also thank M. Mikkelsen and K. Essafi for proof reading the manuscript. Density matrix and state-vector simulations were carried out with \tn{} \cite{robertsTensorNetworkLibraryPhysics2019}. Noise simulations were carried out using \qaer{} \cite{qiskitcontributorsQiskitOpensourceFramework2023}. We used IBM Quantum services to run real device experiments. The views expressed are entirely our own and do not reflect the official policy or position of IBM or the IBM Quantum team.
\end{acknowledgements}

\bibliography{DMERA-writeup}

\appendix

\section{Quantum Circuit and Hardware Details}
\label{app:circuit}
For the diagonalization of the DMERA subsystem we use the parity preserving gate-fabric of the form of \eref{gate-fabric}. This is because although the parity of the subsystem wave-function may differ from that of the total DMERA pure state, the wave-function still remains parity symmetric and the density matrix must be block-diagonal. For an $M=2$ subsystem a single matchgate is enough for the state diagonalization. The diagonalization circuit is a brickwall circuit whose depth in terms of the number of parity preservering gates is $2M-3$.

We used the \qasm{} to estimate the number of shots needed to accurately determine expectation values in a noisy environment. To this end we ran the state diagonalization protocol for a range of shot counts. We determined that 1024 shots per observable evaluation was adequate.
%\begin{figure}
%    \centering
%    \missingfigure{Figure with shot counts indicated}
%    \caption{Results from VQSE with different shot counts}
%    \label{fig:shot_counts}
%\end{figure}

Based on how we choose a contiguous sequence of qubits as a subsystem of DMERA, both the number of qubits, but also the routing of those qubits will change. We will here discuss the consequence of choosing as a subsystem either the left-most or right-most $M$ qubits.

We first note that our chosen gate-fabric $U(\theta_0,\theta_1)$ is not generally symmetric under permutation of its input qubits, and it is also not possible to effectively swap the inputs by applying some transformation to the input angles. However, we may still swap the inputs by applying a transformation to the input angles depending on the ancilla qubit used in that isometry. For the isometry
\begin{align}
    W(\theta_0,\theta_1)
    &= U(\theta_0,\theta_1) \sum_i \ket{i,0}\bra{i,0}\\
    &=
    \begin{bmatrix}
        \cos \theta_1 & 0 \\
        0 & \cos \theta_0 \\
        0 & \sin \theta_0 \\
        \sin \theta_1 & 0 
    \end{bmatrix}
\end{align}
we can apply the transformation $(\theta_0, \theta_1) \rightarrow (\theta_0 - \pi/2, \theta_1)$ to effectively swap the input qubits
\begin{align}
    W(\theta_0 - \pi/2,\theta_1)
    &=
    \begin{bmatrix}
        \cos \theta_1 & 0 \\
        0 & -\sin \theta_0 \\
        0 & \cos \theta_0 \\
        \sin \theta_1 & 0 
    \end{bmatrix} \\
    &= U(\theta_0,\theta_1) \sum_i \ket{0,i}\bra{0,i}.
\end{align}
Similarly, with $\ket{1}$ as the input ancilla qubit, the needed transformation is $(\theta_0, \theta_1) \rightarrow (\theta_0 + \pi/2, \theta_1)$. This allows a swap-gate free transpilation of DMERA for several parameters even with linear device connectivity. In particular if the system is partitioned from the left, with $M=2$ regardless of the depth of DMERA no swap gates are needed to encode the causal cone of its subsystem. For larger $M$ or subsystems that are partitioned from the right, more connections between qubits are needed to avoid swap gates.

\end{document}